\documentstyle[preprint,aps]{revtex}

\begin{document}
\title{Two families of superintegrable and isospectral potentials in two dimensions}
\author{B. Demircio\u{g}lu$^{1}$\thanks{%
Electronic mail: demircio@science.ankara.edu.tr}, \c{S}. Kuru$^{1}$\thanks{%
Electronic mail: kuru@science.ankara.edu.tr}, M. \"{O}nder$^{2}$\thanks{%
Electronic mail: onder@hacettepe.edu.tr}, and A. Ver\c{c}in$^{1}$\thanks{%
Electronic mail: vercin@science.ankara.edu.tr}}
\address{$^{1}$Department of Physics, Ankara University, Faculty of
Sciences,
06100 Tando\u{g}an-Ankara Turkey\\
$^{2}$Department of Physics Engineering, Hacettepe
University, 06532, Beytepe-Ankara, Turkey}
\maketitle

\begin{abstract}
As an extension of the intertwining operator idea, an algebraic method which
provides a link between supersymmetric quantum mechanics and quantum
(super)integrability is introduced. By realization of the method in two
dimensions, two infinite families of superintegrable and isospectral
stationary potentials are generated. The method makes it possible to perform
Darboux transformations in such a way that, in addition to the isospectral
property, they acquire the superintegrability preserving property. Symmetry
generators are second and fourth order in derivatives and all potentials are
isospectral with one of the Smorodinsky-Winternitz potentials. Explicit
expressions of the potentials, their dynamical symmetry generators and the
algebra they obey as well as their degenerate spectra and corresponding
normalizable states are presented.
\end{abstract}

\tightenlines
\section{Introduction}
A Hamiltonian system of $N$ degrees of freedom is said to be completely
integrable, in the Liouville-Arnold sense, if it possesses functionally
independent globally defined and single-valued $N$ integrals of motion in
involution \cite{Arnold,Goldstein}. It is called to be superintegrable if it
admits more than $N$ integrals of motion. Not all the integrals of
superintegrable system can be in involution, but they must be functionally
independent otherwise the extra invariants are trivial. In analogy to the
classical mechanics, a quantum mechanical system described in $N$%
-dimensional ($N$D) Euclidean space by a stationary Hamiltonian operator $H$
is called to be completely integrable if there exists a set of $N-1$
(together with $H,N$) algebraically independent linear operators $X_{i},i=1,
2, ..., N-1$ commuting with $H$ and among each other \cite
{Fris,Makarov,Wojciechowski,Evans1,Evans2,Grosche,Ranada,Sheftel,Tempesta}.
If there exist $k$ additional operators $Y_{j},j=1, 2, ..., k$ where $%
0<k\leq N-1$, commuting with $H$ it is said to be superintegrable. The
superintegrability is said to be minimal if $k=1$ and maximal if $k=N-1$.

Classical and quantum mechanical examples of the maximally superintegrable
systems for any finite $N$ are; the Kepler-Coulomb problem, the harmonic
oscillator with rational frequency ratio, the Calogero-Moser system in a
harmonic well and the Winternitz (or, Smorodinsky-Winternitz) system. The
first two are known, for $N=3$, since the time of Laplace and the
superintegrability of the last two systems were established for the first
time, respectively, by Wojciechowski \cite{Wojciechowski} and by Evans \cite
{Evans1}. The first systematic search for other possible superintegrable
systems was begun by Winternitz and coworkers. They firstly found four
independent 2D potentials that are separable in more than one coordinate
system \cite{Fris}, and then they extended this to $N=3$ \cite{Makarov}.
This approach is based on two assumptions; (1) Hamiltonians are of potential
form. (2) Integrals of motion are at most quadratic in momenta (or, in
derivatives). The Winternitz program has been completed in Ref.\cite{Evans2}
where a complete list consisting, up to the equivalence of linear
transformations, thirteen different 3D potentials with four or five
independent integrals of motion is given. Winternitz potentials have also
been considered by different formulations such as path integral formulation 
\cite{Grosche}, Lagrangian formalism \cite{Ranada} and evolutionary vector
fields formalism \cite{Sheftel}.

In this paper we report an infinite family of 2D potentials which are not
only superintegrable, but at the same time isospectral. We shall give
explicit expressions of the potentials, their dynamical symmetry generators
and the algebra they obey as well as their degenerate spectrum and
corresponding normalizable states. We achieve this goal by following an
algebraic method which is based on and, in fact, is an extension of
intertwining operator idea. This is closely connected with the
supersymmetric (SUSY) methods such as the Darboux transformation, and
Schr\"{o}dinger factorization which deal with pairs of Hamiltonians having
the same energy spectra but different eigenstates \cite
{Junker,Cooper,Matveev,Aratyn}. It turns out that each member of this
infinite family is a triplet of potentials one of which is the same for
entire family and the other two change from member to member. Hence, we
have, in fact, two different infinite families of superintegrable and
isospectral potentials. The fixed potential turns out to be one of $2D$
Winternitz potentials and determines the spectra of both families and the
other two are intertwined to it by Darboux type transformations. The
generators of these transformations depend on eigenfunctions of two
associated solvable 1D problems that result from the separation of the
Winternitz potential in different coordinates. We should emphasize that our
approach makes it possible to apply Darboux transformations simultaneously
to potential and to its symmetry generators in such a way that
superintegrability property is preserved.

Formal aspects of our method together with a brief review of the main points
of the intertwining operator idea will be given in the next section. Secs.
III-V are devoted to explicit realization of our method. In Sec. IV we
present the most general form of 2D integrable and isospectral potentials in
the plane polar coordinates. Two subfamilies of superintegrable and
isospectral potentials and then their general forms are presented in Secs.
VI and VII. Sec. VIII contains a review of bound states of the associated 1D
problems and the above mentioned Winternitz potential. After investigating
the symmetry generators and their algebra in Sec. IX, the normalizable
states of the generated superintegrable and isospectral potentials are given
in Sec. X.

\section{Multiple Intertwining Method}

The object of the intertwining method is to construct a linear differential
operator ${\cal L}$ which intertwines two Hamiltonian operators $H_{0}$ and $%
H_{1}$ such that ${\cal L}H_{0}=H_{1}{\cal L}$. Two important facts that
immediately follow from this relation are; (i) If $\psi^{0}$ is an
eigenfunction of $H_{0}$ with eigenvalue of $E^{0}$ then $\psi ^{1}={\cal L}%
\psi^{0}$ is an (unnormalized) eigenfunction of $H_{1}$ with the same
eigenvalue $E^{0}$. (ii) When $H_{0}$ and $H_{1}$ are self-adjoint (on some
common function space) ${\cal L}^{\dagger }$ intertwines in the other
direction $H_{0}{\cal L}^{\dagger }={\cal L}^{\dagger }H_{1}$ and this in
turn implies that $[H_{0}, {\cal L}^{\dagger }{\cal L}]=0=[{\cal L}{\cal L}%
^{\dagger },H_{1}]$, where $^{\dagger }$ and $[,]$ stand for Hermitian
conjugation and commutator. The first property shows that ${\cal L}$
transforms one solvable problem into another, and the second one means that
two hidden dynamical symmetries of $H_{0}$ and $H_{1}$ are immediately
constructed in terms of ${\cal L}$. These are dimension and form independent
general properties of this method \cite{Anderson2,Cannata}. In the context
of 1D systems where ${\cal L}$ is taken to be the first order differential
operator and Hamiltonians are of the potential forms two additional
properties arise; (i) Every eigenfunction of $H_{0}$ (without regard to
boundary conditions or normalizability) can be used to generate a
transformation to a new solvable problem. (ii) A direct connection to a SUSY
algebra can be established \cite{Anderson3}. The first property is a
manifestation of the celebrated Darboux transformation and its
generalization (Crum transformation). The second property enables us to
express in a compact algebraic form of the spectral equivalence of the
intertwined systems.

Now suppose that there are three self-adjoint Hamiltonian operators $%
H_{0},H_{1},H_{2}$ which are intertwined as 
\begin{eqnarray}
{\cal L}_{10}H_{0}=H_{1}{\cal L}_{10},\quad {\cal L}_{21}H_{1}=H_{2}{\cal L}%
_{21}.
\end{eqnarray}
The subscripts of the intertwining operators are used to distinguish them
and to denote the intertwined Hamiltonians. Eqs. (1) immediately imply that $%
{\cal L}_{20}\equiv {\cal L}_{21}{\cal L}_{10}$ will intertwine $H_{0}$ and $%
H_{2}$ as follows, 
\begin{eqnarray}
{\cal L}_{20}H_{0}=H_{2}{\cal L}_{20}.
\end{eqnarray}
Eqs. (1) and (2) can be unified into the following diagram 
\begin{eqnarray}
\left. 
\begin{array}{ccc}
H_{0} & \longrightarrow & H_{1} \\ 
& \searrow & \downarrow \\ 
&  & H_{2}
\end{array}
\right.
\end{eqnarray}
which must be understood in the sense described by (1) and (2).

Adjoints of (1) and (2) yield 
\begin{eqnarray}
{\cal L}_{21}^{\dagger }H_{2}=H_{1}{\cal L}_{21}^{\dagger },\quad {\cal L}%
_{10}^{\dagger }H_{1}=H_{0}{\cal L}_{10}^{\dagger },\quad {\cal L}%
_{20}^{\dagger }H_{2}=H_{0}{\cal L}_{20}^{\dagger }.
\end{eqnarray}
That is, the adjoints of the intertwining operators will intertwine in the
reverse directions and this can be represented by a diagram the same as (3)
with reversed directions of arrows. Making use of (1-4) it is easy to show
that each of $H_{0},H_{1},H_{2}$ has two dynamical symmetry generators
respectively given by; 
\begin{eqnarray}
X_{0}={\cal L}_{10}^{\dagger }{\cal L}_{10} &&,\qquad Y_{0}={\cal L}%
_{20}^{\dagger }{\cal L}_{20},  \nonumber \\
X_{1}={\cal L}_{10}{\cal L}_{10}^{\dagger } &&,\qquad Y_{1}={\cal L}%
_{21}^{\dagger }{\cal L}_{21}, \\
X_{2}={\cal L}_{21}{\cal L}_{21}^{\dagger } &&,\qquad Y_{2}={\cal L}_{20}%
{\cal L}_{20}^{\dagger }.  \nonumber
\end{eqnarray}
The subscripts of $X_{j},Y_{j}$ indicate the Hamiltonians they belong to.
Throughout this paper we assume that the domains of definition of
Hamiltonians and intertwining operators are some linear subspaces of a
common Hilbert space ${\cal H}=L^{2}(\Omega )$ with the standard
sesquilinear inner product. $L^{2}(\Omega )$ is the space of all
square-integrable functions (and distributions) defined on a subspace $%
\Omega $ of $N$D Euclidean space $R^{N}$ \cite{Richtmyer,Kato,Reed}.

For all $N\geq 2$, the diagram (3) implies a triplet of isospectral
Hamiltonians such that each has two dynamical symmetries. By construction,
all the symmetry operators obtained in this manner will be factorized, and
have even orders depending on the order of intertwining operators. They will
be of the same order only for $H_{1}$. According to the von Neumann theorem
(see Ref.\cite{Richtmyer}, pp.141 and Ref.\cite{Kato}, pp.275) ${\cal L}_{ij}%
{\cal L}^{\dagger}_{ij}$ ( and ${\cal L}^{\dagger}_{ij}{\cal L}_{ij}$) are
self-adjoint and nonnegative if ${\cal L}_{ij}$ are closed with dense
domains of definition. Otherwise there may exist states in which they have
negative expectation values (see Sec. X). If ${\cal L}_{10}$ and ${\cal L}%
_{21}$ are taken to be algebraically independent, the independence of $%
X_{i},Y_{i}$ pairs will be guaranteed from the outset. Extensions of these
ideas to higher dimensions will be, generically, called multiple
intertwining method. A simple observation that this work initiated from is
that, in the particular case of $N=2$ the diagram (3) guarantees the
superintegrability of the three Hamiltonians. In the case of $N=3$ such a
diagram will imply, provided that symmetry generators are commutative, the
integrability of the potentials.

The rest of the paper is devoted to explicit realization of these formal
observations for 2D systems. Firstly we will determine the most general form
of the potentials and the first order intertwining operator for two
Hamiltonians. We then construct the intertwinings $H_{0}\rightarrow H_{1}$
and $H_{1}\rightarrow H_{2}$ by special forms of the intertwining operator.
We end this section by explaining our use of the adjective ``isospectral''.
Two Hamiltonians are said to be isospectral if they have the same eigenvalue
spectrum \cite{Anderson3,Deift,Pursey}. In this sense two linearly
intertwined Hamiltonians are always formally isospectral except the
eigenvalues corresponding to the kernel of the intertwining operator. Even
for these exceptional cases one can construct eigenfunctions corresponding
to these eigenvalues at least for 1D and 2D systems by appealing to the
Liouville formula and its 2D version \cite{Matveev}. However, due to
physical requirements, in the case of the bound states mainly due to
normalizability conditions, some eigenvalues of one of the partner
potentials are to be discarded. For higher dimensional systems also the
degree of degeneracy of a common eigenvalue may be different (see Sec. X).
These will just mean that a finite number of eigenvalues are to be
disregarded for they are not physically admissible.

\section{Intertwining in Two Dimensions}

We start by considering a pair of 2D one particle systems characterized by
the Hamiltonian operators of potential form, 
\begin{equation}
H_{i}=-\nabla ^{2}+V_{i},\quad H_{f}=-\nabla ^{2}+V_{f},
\end{equation}
where the potentials $V_{i},V_{f}$ (and eigenvalues of $H_{i},H_{f}$) are
expressed in terms of $2m/\hbar ^{2}$ and 
\begin{eqnarray}
\nabla ^{2}=\partial _{r}^{2}+\frac{1}{r}\partial _{r}+\frac{1}{r^{2}}%
\partial^{2}_{\theta }  \nonumber
\end{eqnarray}
is the Laplace operator in the plane polar coordinates $(r,\theta )$. $m$ is
the mass of the particle and $\hbar $ denotes the Planck constant. Here and
hereafter we use the notation $\partial _{x}$ for partial derivative $%
\partial /\partial x$ and the subindexes $``i",``f"$ as the shorthands for
the ``initial'' and ``final''. We suppose that the Hamiltonians are
intertwined by 
\begin{equation}
{\cal L}_{fi}H_{i}=H_{f}{\cal L}_{fi}
\end{equation}
and propose the ansatz that ${\cal L}_{fi}$ is the most general first order
linear operator 
\begin{equation}
{\cal L}_{fi}=L_{0}+L_{d}=L_{0}+L_{1}\partial _{r}+L_{2}\partial _{\theta },
\end{equation}
where $L_{d}=L_{1}\partial _{r}+L_{2}\partial _{\theta }$ will be referred
to as the differential part of ${\cal L}_{fi}$. The potentials and $%
L_{0},L_{1},L_{2}$ are some real functions of $(r,\theta )$ which are to be
determined from consistency equations of the intertwining relation (7).

In view of (6) and (8) the relation (7) explicitly reads as 
\begin{equation}
\lbrack \nabla ^{2},L_{d}]=-[\nabla ^{2},L_{0}]+[V_{i},L_{d}]+P{\cal L}_{fi},
\end{equation}
where $P=V_{f}-V_{i}$. The second order derivatives come, together with some
first order derivatives, only from 
\begin{eqnarray}
\lbrack \nabla ^{2},L_{d}] &=&(\nabla ^{2}L_{1}+\frac{1}{r^{2}}%
L_{1})\partial _{r}+(\nabla ^{2}L_{2})\partial _{\theta }+  \nonumber \\
&&2(\frac{1}{r^{2}}\partial _{\theta }L_{1}+\partial _{r}L_{2})\partial
_{\theta }\partial _{r}+2(\partial _{r}L_{1})\partial _{r}^{2}+\frac{2}{r^{3}%
}(L_{1}+r\partial _{\theta }L_{2})\partial _{\theta }^{2},
\end{eqnarray}
and by setting their coefficients to zero we obtain; 
\begin{eqnarray}
\partial _{\theta }L_{1}+r^{2}\partial _{r}L_{2}=0,\quad \partial
_{r}L_{1}=0,\qquad L_{1}+r\partial _{\theta }L_{2}=0.  \nonumber
\end{eqnarray}
It is straightforward to show that the general solutions of these equations
are 
\begin{eqnarray}
L_{1}=A\sin (\theta +\phi ),\qquad L_{2}=B+\frac{A}{r}\cos (\theta +\phi ),
\end{eqnarray}
where $A,B$ and $\phi $ are integration constants. Since $\nabla
^{2}L_{1}=-L_{1}/r^{2}$, and $\nabla ^{2}L_{2}=0$, we have from (10) $%
[\nabla ^{2},L_{d}]=0$. As a result of this the relation (9) simplifies to 
\begin{equation}
\lbrack \nabla ^{2},L_{0}]=-L_{1}\partial _{r}V_{i}-L_{2}\partial _{\theta
}V_{i}+P(L_{0}+L_{d}).
\end{equation}

By substituting 
\begin{eqnarray}
\lbrack \nabla ^{2},L_{0}]=\nabla ^{2}L_{0}+ 2(\partial _{r}L_{0})\partial
_{r}+\frac{2}{r^{2}}(\partial _{\theta }L_{0})\partial _{\theta },  \nonumber
\end{eqnarray}
into (12), and then by equating the coefficients of the first and zeroth
powers of derivatives we obtain 
\begin{eqnarray}
2\partial _{r}L_{0} &=&PL_{1}, \\
2\partial _{\theta }L_{0} &=&r^{2}PL_{2}, \\
(-\nabla ^{2}+P)L_{0} &=&L_{1}\partial _{r}V_{i}+L_{2}\partial _{\theta
}V_{i}.
\end{eqnarray}
These three partial differential equations, the first two of which are
linear and the third is nonlinear, constitute a reduced form of the
consistency conditions for three unknown functions $L_{0},V_{i}$ and $V_{f}$.

\section{General Form of 2D Integrable Isospectral Potentials in Polar
Coordinates}

Eqs. (11), (13-14) and the compatibility condition $\partial _{r}\partial
_{\theta }L_{0}=\partial _{\theta }\partial _{r}L_{0}$ imply that 
\[
2\nabla ^{2}L_{0}=L_{d}P,\quad ZL_{0}=0,\quad ZP=2BrP, 
\]
where $Z=L_{1}\partial _{\theta }-r^{2}L_{2}\partial _{r}$. From the second
and third of these equations (or, from (13) and (14)) we have $L_{0}=f(w)$,
and $P=-2A^{2}f^{\prime }(w)/r^{2}L_{1}^{2}$, where $f$ is an arbitrary
function of 
\[
w=B\cot (\theta +\phi )+\frac{A}{r\sin (\theta +\phi )}. 
\]
Prime stands for derivative with respect to the argument and when there is
no risk of confusion the argument will be suppressed. By combining $2\nabla
^{2}L_{0}=L_{d}P$ with (15) and using the found $L_{0}$ and $P$ we obtain an
inhomogeneous equation from the general solution of which the general form
of potentials are found to be 
\begin{eqnarray}
V_{i}=h(\kappa )+\frac{{\cal V}_{-}(w)}{\kappa ^{2}},\quad V_{f}=h(\kappa )+%
\frac{{\cal V}_{+}(w)}{\kappa ^{2}}.
\end{eqnarray}
Here $h$ is an arbitrary function of $\kappa =[A^{2}+B^{2}r^{2}+2ABr\cos
(\theta +\phi )]^{1/2}$ such that $L_{d}h=0$ and 
\begin{eqnarray}
{\cal V}_{\pm }(w)=f^{2}(w)\mp (w^{2}+B^{2})f^{\prime }(w).
\end{eqnarray}
Eqs. (16) represent the most general form of 2D integrable and isospectral
potentials in polar coordinates.

Let us define the operators 
\begin{eqnarray}
T_{1}=\cos\theta \partial_{r} -\frac{1}{r}\sin\theta\partial_{\theta},\quad
T_{2}=\sin\theta \partial_{r}+ \frac{1}{r}\cos\theta\partial_{\theta},\quad
J=\partial_{\theta},
\end{eqnarray}
which close in the defining relations of the Euclidean Lie algebra $e(2)$ in
two dimensions 
\begin{eqnarray}
[J, T_{1}]=-T_{2},\quad [J, T_{2}]=T_{1},\quad [T_{1}, T_{2}]=0.
\end{eqnarray}
Now $L_{d}$ can be rewritten as 
\begin{equation}
L_{d}= A\sin\phi T_{1}+A\cos\phi T_{2}+BJ,
\end{equation}
which shows that the differential part of ${\cal L}_{fi}$ is an element of $%
e(2)$. In terms of the Cartesian coordinates $x=r\cos\theta, y=r\sin\theta $
we have $T_{1}=\partial_{x}, T_{2}= \partial_{y},
J=x\partial_{y}-y\partial_{x}$ and $T_{i}^{\dagger}=-T_{i}, J^{\dagger}=-J$.
These relations can also be verified from (18) by noting that $%
(\partial_{r})^{\dagger}=-(r^{-1}+\partial_{r}),
(\partial_{\theta})^{\dagger}=-\partial_{\theta}$. Now from (5) and (17) the
symmetry generators of $H_{i}$ and $H_{f}$ are 
\begin{eqnarray}
{\cal L}_{fi}^{\dagger }{\cal L}_{fi}={\cal V}_{-}-L^{2}_{d},\quad {\cal L}%
_{fi}{\cal L}_{fi}^{\dagger } ={\cal V}_{+}-L^{2}_{d},  \nonumber
\end{eqnarray}
where $L_{d}^{2}$ is at most quadratic operator in generators of $e(2)$.

\section{Construction of the intertwining operators}

We shall construct the legs of the diagram (3) by adopting particular forms
of (20) as the differential parts of ${\cal L}_{10}$ and ${\cal L}_{21}$. In
doing that we shall make use of the orbit structure of $e(2)$ under the
adjoint action of the Euclidean group $E(2)$ in two dimensions \cite{Miller}.

Under a unitary similarity transformation, generated by 
\begin{eqnarray}
U=e^{a_{0}J}e^{a_{1}T_{1}+a_{2}T_{2}},\quad
U^{\dagger}=U^{-1}=e^{-(a_{1}T_{1}+a_{2}T_{2})}e^{-a_{0}J},
\end{eqnarray}
where $a_{i}$'s are real parameters and $U^{-1}$ stands for the inverse of $%
U\in E(2)$, the relation (7) transforms into $\bar{{\cal L}}_{fi}\bar{H}_{i}=%
\bar{H}_{f}\bar{{\cal L}}_{fi}$, where $\bar{X}=UXU^{\dagger}$. Since $%
\nabla^{2}=T^{2}_{1}+T^{2}_{2}$ is the Casimir invariant of $e(2)$, only $%
V_{i}, V_{f}$ and ${\cal L}_{fi}$ will change under this $E(2)$ action. Now
suppose that $L_{d}$ is of the form (20). Making use of the well known
operator identity 
\begin{eqnarray}
e^{bK}Me^{-bK}=M+b[K, M]+\frac{b^{2}}{2!}[K,[K, M]]+\cdots,  \nonumber
\end{eqnarray}
where $b$ is a constant and $K, M$ are two arbitrary operators, one can
easily show that 
\begin{eqnarray}
\bar{L}_{d}=BJ+ e^{a_{0}J}[T_{1}(A\sin\phi - a_{2}B)+T_{2}(A\cos\phi +
a_{1}B)]e^{-a_{0}J}.  \nonumber
\end{eqnarray}
Hence, if $B\neq 0$ we can take $\bar{L}_{d}=BJ$ by choosing $%
a_{1}=-A\cos\phi/B, a_{2}=A\sin\phi/B$. On the other hand, if $B=0, A\neq 0$
we get $\bar{L}_{d}=AT_{1}$ (or, $\bar{L}_{d}=AT_{2}$) for the choice $%
a_{0}=\phi$ (or, $a_{0}=-\phi$). Therefore, under the adjoint action of $%
E(2) $, $e(2)$ has two orbits represented by $J$ and $T_{2}$. Since $L_{d}$
and $cL_{d}$ belong to the same orbit for $c\neq 0$, we can choose $L_{d}=J$
for ${\cal L}_{10}$ and $L_{d}=T_{2}$ for ${\cal L}_{21} $. In such a case
the potentials and $L_{0}$ will be specified up to the adjoint action of $%
E(2)$.

For the first leg $H_{0}\rightarrow H_{1}$ of (3) we take $A=0, B=1$ in Eq.
(11) and redefine the Hamiltonians as $H_{i}=H_{0}$ and $H_{f}=H_{1}$. Hence 
$L_{1}=0, L_{2}=1$ and Eqs. (13-14) imply that $L_{0}=f(\theta)$ and 
\begin{eqnarray}
{\cal L}_{10}=f(\theta)+\partial_{\theta}, \quad P=V_{1}-V_{0}= \frac{2}{%
r^{2}}f^{\prime}(\theta),
\end{eqnarray}
where $f$ is an arbitrary differentiable function of $\theta$. Noting that $%
\nabla^{2}L_{0}=f^{\prime \prime}(\theta)/r^{2}$ we obtain from (15) and
(22) 
\begin{equation}
V_{0}=h(r)+\frac{V_{-}(\theta)}{r^{2}},\quad V_{1}=h(r)+\frac{V_{+}(\theta)}{%
r^{2}},
\end{equation}
where $h$ is an arbitrary differentiable function of $r$ and 
\begin{equation}
V_{\pm}(\theta)= f^{2}(\theta)\pm f^{\prime}(\theta).
\end{equation}
As a result the first $H_{0}\rightarrow H_{1}$ leg of the diagram (3) has
been constructed.

For the second leg we take $B=0, A=1$, fix the form of $H_{1}$ and denote it
as $H_{i}=H_{1}$. We then look for $H_{f}=H_{2}$ such that ${\cal L}%
_{21}H_{1}=H_{2}{\cal L}_{21}$ and ${\cal L}_{21}=L_{0}+\sin\phi
T_{1}+\cos\phi T_{2} $. In that case from Eqs. (13-14) we get $L_{0}=g(u)$
and 
\begin{eqnarray}
{\cal L}_{21}&=&g(u)+\sin(\theta+\phi)\partial_{r} +\frac{1}{r}%
\cos(\theta+\phi)\partial_{\theta}, \\
P&=&V_{2}-V_{1}=2g^{\prime}(u),
\end{eqnarray}
where $g$ is an arbitrary differentiable function of $u=r\sin(\theta+\phi)$.
It only remains to solve the nonlinear equation (15) which now takes the
form 
\begin{equation}
\partial_{u}[g^{2}(u)-g^{\prime}(u)]=\sin(\theta+\phi)h^{\prime}(r) +\frac{1%
}{r^{3}}[
\cos(\theta+\phi)V_{+}^{\prime}(\theta)-2\sin(\theta+\phi)V_{+}(\theta)],
\end{equation}
where we have made use of $\nabla ^{2}L_{0}=g^{\prime \prime}(u)$ and of the
second equation of (23). Note that we could have chosen $\phi=0$, but since
it costs almost nothing we keep $\phi$ in our formulae in order to see that
action of $E(2)$.

Since it further restricts the three arbitrary functions specifying the
potentials, Eq. (27) is the main equation which determines the final form of
the potentials. As a consistency condition the right hand side of Eq. (27)
must be only a function of $u$. Nevertheless this requirement provides us
with many possibilities for $f, g$ and $h$, which are investigated in the
next two sections. Note that for any solutions of Eq. (27) the potentials
will be connected to each other as follows: 
\begin{eqnarray}
V_{0}=V_{1}-\frac{2}{r^{2}}f^{\prime}(\theta), \quad
V_{2}=V_{1}+2g^{\prime}(u),\quad V_{0}=V_{2}-2[g^{\prime}(u)+\frac{%
f^{\prime}(\theta)}{r^{2}}].
\end{eqnarray}

\section{Two Subfamilies of Potentials}

We construct the simplest family of potentials by taking, in (24) and (27) $%
h=(\lambda _{1}/r^{2})+a,V_{+}=-\lambda _{1}$. These lead us to 
\begin{equation}
f^{2}+f^{\prime }=-\lambda _{1},\qquad g^{2}-g^{\prime }=-\lambda _{2},
\end{equation}
where $a,\lambda _{1},\lambda _{2}$ are some arbitrary constants. Then, by
Eqs. (23-24) and (26), we obtain 
\begin{equation}
V_{0}=\frac{2}{r^{2}}(f^{2}+\lambda _{1})+a,\qquad V_{1}=a,\qquad
V_{2}=2(g^{2}+\lambda _{2})+a.
\end{equation}
The general solution of $g^{2}-g^{\prime }=-\lambda _{2}$ is 
\[
g=\left\{ 
\begin{array}{cc}
\lambda _{2}^{1/2}\tan (\lambda _{2}^{1/2}u+a_{1}); & \quad for\quad \lambda
_{2}>0, \\ 
-\frac{1}{u+a_{1}}; & \quad for\quad \lambda _{2}=0, \\ 
(-\lambda _{2})^{1/2}\tanh [(-\lambda _{2})^{1/2}u+a_{1}]; & \quad for\quad
\lambda _{2}<0,
\end{array}
\right. 
\]
where $a_{1}$ is a constant. The solution of $f^{2}+f^{\prime }=-\lambda
_{1} $ can be directly read from the above relation after the replacement $%
(g,u,\lambda _{2})\rightarrow (f,-\theta ,\lambda _{1})$.

An important point is that, by the usual linearization of the Riccati
equation, if we substitute 
\begin{eqnarray}
f(\theta)=\frac{\psi^{\prime}(\theta)}{\psi(\theta)}, \quad g(u)=-\frac{%
\Psi^{\prime}(u)}{\Psi(u)},
\end{eqnarray}
into (29) we arrive at two 1D Schr\"{o}dinger equations 
\begin{eqnarray}
-\psi^{\prime\prime}(\theta)=\lambda_{1} \psi(\theta),\quad -\Psi^{\prime
\prime}(u)=\lambda_{2} \Psi(u).
\end{eqnarray}
While the second one can be considered as a free motion, this is not the
case for the first since $0\leq \theta <2\pi$. An appealing case is to
consider one, or, both of them as infinite square-well problem. Normalized
eigenfunctions subjected to boundary conditions, say, $\psi(0)=0=\psi(2\pi)$
and corresponding eigenvalues are 
\begin{eqnarray}
\psi_{k}(\theta)=\pi^{-1/2}\sin(\frac{1}{2}k\theta ),\quad \lambda_{1,k}=%
\frac{k^{2}}{4}, \quad k=1, 2,...
\end{eqnarray}
Hence $f_{k}=(k/2)\cot(k\theta/2)$ and by virtue of Eqs. (22) and (24) we
have 
\begin{eqnarray}
V^{(k)}_{0}=\frac{k^{2}}{2r^{2}\sin^{2}\frac{1}{2}k\theta}+a ,\quad {\cal L}%
^{(k)}_{10}=\frac{k}{2}\cot(\frac{1}{2}k\theta)+\partial_{\theta}.
\end{eqnarray}
To distinguish the resulting potentials, corresponding intertwining
operators and the parameter $\lambda_{1}$ we have labelled them by the
quantum number $k$. The $u$-problem can be treated in a similar way. In any
case, the potentials and transformations among them are generated by
solutions of these two auxiliary 1D problems. The existence of $V_{1}=a $
explicitly shows that the member potentials are isospectral to a 2D free
motion. As a result we have found a five parameter ($a, a_{1}, \lambda_{1},
\lambda_{2}, \phi$) family of 2D potentials that are generated, in a
nontrivial way, by two 1D problems.

We specify a second subfamily of potentials by taking, in (23-24) and (27) 
\begin{equation}
h=\frac{\lambda_{1} }{r^{2}}+\frac{1}{2}\alpha r^{2}+a,
\end{equation}
and $V_{+}=-\lambda_{1}$. These lead us to the same equation as in (29) for $%
f$ and to the Riccati's equation 
\begin{equation}
g^{2}-g^{\prime} -\frac{1}{2}\alpha u^{2} +\lambda_{2}=0,
\end{equation}
for $g$. By Eqs. (23-24) and (26) the member potentials are found to be 
\begin{eqnarray}
V_{0}&=& \frac{1}{2}\alpha r^{2}+\frac{2(f^{2}+\lambda_{1} )}{r^{2}}+a, 
\nonumber \\
V_{1}&=& \frac{1}{2}\alpha r^{2}+a, \\
V_{2}&=& \frac{1}{2}\alpha
r^{2}\cos2(\theta+\phi)+2g^{2}(u)+(a+2\lambda_{2}),  \nonumber
\end{eqnarray}
where $g$ is any solution of (36) and $f$ is any solution of $%
f^{2}+f^{\prime}=-\lambda_{1} $.

Now the ansatz (31) for $g$ transforms (36) into 
\begin{equation}
-\Psi ^{\prime \prime }(u)+\frac{1}{2}\alpha u^{2}\Psi (u)=\lambda _{2}\Psi
(u),
\end{equation}
which is the well known Schr\"{o}dinger equation for the 1D harmonic
oscillator. In that case the entire family will have 2D isotropic harmonic
oscillator spectrum given by the eigenvalues 
\begin{eqnarray}
E_{\ell }^{(1)}=\hbar \omega (\ell +1),\quad \ell =0,1,2,...,
\end{eqnarray}
which are $\ell +1$ times degenerate for a given $\ell $. For concrete
examples we recall the normalized eigenfunctions and corresponding
eigenvalues of the 1D harmonic oscillator: 
\begin{eqnarray}
\Psi _{n}(u)=N_{n}e^{-\beta ^{2}u^{2}/2}H_{n}(\beta u),\quad E_{n}=\frac{%
\hbar ^{2}}{2m}\lambda _{2,n}=\hbar \omega (n+\frac{1}{2}),\quad
n=0,1,2,\dots ,
\end{eqnarray}
where $N_{n}$ is the normalization constant, $H_{n}$ denote the Hermite
polynomials and 
\begin{eqnarray}
\beta =(\frac{m\omega }{\hbar })^{1/2}=(\frac{\alpha }{2})^{1/4},\quad
N_{n}=(\frac{\beta }{\pi ^{1/2}2^{n}n!})^{1/2}.
\end{eqnarray}
In writing Eqs. (39-41) we have restored $2m/\hbar ^{2}$ into our notation
in which the dimension of $\beta $ is $(length)^{-1}$. Like $\lambda _{2}$,
also $V_{2},{\cal L}_{21}$ and the function $g$ must be labelled by the
quantum number $n$: 
\begin{eqnarray}
g_{n}(u) &=&-\frac{\Psi _{n}^{\prime }(u)}{\Psi _{n}(u)}=\beta
^{2}u-\partial _{u}\ln [H_{n}(\beta u)],  \nonumber \\
V_{2}^{(n)} &=&\beta ^{4}r^{2}\cos 2(\theta +\phi )+2g_{n}^{2}(u)+a+4\beta
^{2}(n+\frac{1}{2}), \\
{\cal L}_{21}^{(n)} &=&g_{n}(u)+\sin (\theta +\phi )\partial _{r}+\frac{1}{r}%
\cos (\theta +\phi )\partial _{\theta }.  \nonumber
\end{eqnarray}
For the first three Hermite polynomials $%
H_{0}(x)=1,H_{1}(x)=2x,H_{2}(x)=4x^{2}-2$ we have 
\[
g_{0}=\beta ^{2}u,\quad g_{1}=\beta ^{2}u-\frac{1}{u},\quad g_{2}=\beta ^{2}%
\frac{2\beta ^{2}u^{2}-5}{2\beta ^{2}u^{2}-1}u. 
\]
Considering the $f$-problem as above ${\cal L}_{10}^{(k)}$ is given by (34)
and $V_{0}$ is 
\begin{equation}
V_{0}^{(k)}=\frac{1}{2}\alpha r^{2}+\frac{k^{2}}{2r^{2}\sin ^{2}(\frac{1}{2}%
k\theta )}+a.
\end{equation}

\section{General Form of the Potentials}

Returning the general discussion of Sec. V, the most general potentials are
obtained by choosing, in Eq. (27), $h$ as in (35) and by postulating the
equation 
\begin{equation}
\cos (\theta +\phi )V_{+}^{\prime }(\theta )-2\sin (\theta +\phi
)V_{+}(\theta )=2\lambda_{1}\sin (\theta +\phi )-\frac{2c}{\sin ^{3}(\theta
+\phi )},
\end{equation}
for $V_{+}$. It is not hard to check that (35) and (44) are the most general
relations which make the right hand side of Eq. (27) only a function of the $%
u$ variable. The general solution of Eq. (44) is 
\begin{equation}
V_{+} =f^{2}(\theta)+f^{\prime}(\theta) =\frac{b}{\cos^{2}(\theta +\phi )}+%
\frac{c}{\sin ^{2}(\theta +\phi )}-\lambda_{1},
\end{equation}
where $\lambda_{1}, b $ and $c$ are some constants. When (35) and (44) are
inserted into (27) we obtain a new Riccati's equation for $g(u)$ 
\begin{equation}
g^{2}-g^{\prime }=\frac{1}{2}\alpha u^{2}+\frac{c}{u^{2}}-\lambda_{2}.
\end{equation}
By virtue of (23), (28), (35), (45) and (46) the corresponding potentials
can be written as 
\begin{eqnarray}
V_{0} &=&\frac{1}{2}\alpha r^{2}-\frac{1}{r^{2}} [\frac{b }{\cos ^{2}(\theta
+\phi )}+\frac{c}{\sin ^{2}(\theta +\phi )}] +\frac{2(f^{2}+\lambda _{1})}{%
r^{2}}+a,  \nonumber \\
V_{1} &=&\frac{1}{2}\alpha r^{2}+\frac{1}{r^{2}} [\frac{b }{\cos ^{2}(\theta
+\phi )}+\frac{c}{\sin ^{2}(\theta +\phi )}]+a, \\
V_{2} &=&\frac{1}{2}\alpha r^{2}\cos 2(\theta +\phi )+\frac{1}{r^{2}} [\frac{%
b }{\cos ^{2}(\theta +\phi )}-\frac{c}{\sin ^{2}(\theta +\phi )}]
+2(g^{2}+\lambda _{2}+\frac{a}{2}).  \nonumber
\end{eqnarray}

$V_{1}$ is immediately recognized as one of 2D Smorodinsky-Winternitz
potentials which accepts separation of variables in the Cartesian, polar and
elliptic coordinates. Being fixed in the whole family it determines the
structure of spectrum of all potentials. While $V_{0}$ is separable in the
plane polar coordinates $V_{2}$ is separable only in the Cartesian
coordinates. $V_{0}$ and $V_{2}$ represent two families of the
superintegrable and isospectral potentials generated by the functions $f$
and $g$ which are subjected to Eqs. (45) and (46). The normalized
eigenfunctions, corresponding eigenvalues and the symmetry generators will
be the subject of the next three sections.

Having specified the most general forms of the potentials we now show how to
develop a hierarchy of the potentials.

On substituting (31) into (45) and (46) we arrive at the following two 1D
problems 
\begin{equation}
H_{PT}\psi _{k}(\theta )=\lambda _{1, k}\psi _{k}(\theta ),\quad H_{SO}\Psi
_{n}(u)=\lambda _{2, n}\Psi _{n}(u),
\end{equation}
where $k$ and $n$ are possible quantum numbers and 
\begin{eqnarray}
H_{PT} &=&-\frac{d^{2}}{d\theta ^{2}}+V_{PT},\qquad V_{PT}=\frac{b}{\cos
^{2}(\theta +\phi )}+\frac{c}{\sin ^{2}(\theta +\phi )}, \\
H_{SO} &=&-\frac{d^{2}}{du^{2}}+V_{SO},\qquad V_{SO}=\frac{1}{2}\alpha u^{2}+%
\frac{c}{u^{2}}.
\end{eqnarray}
These are the well-known generalized P\"{o}schl-Teller (PT) and singular
oscillator (hence the subscript SO), or the radial oscillator potentials. By
virtue of (28) and (31) the potentials can be rewritten as 
\begin{eqnarray}
V_{0}^{(k)}=V_{1}-\frac{2}{r^{2}}\partial _{\theta }^{2}\ln \psi _{k
}(\theta ),\quad V_{2}^{(n)}=V_{1}-2\partial _{u}^{2}\ln \Psi _{n}(u).
\end{eqnarray}
Here and here after we label the potentials by the quantum numbers of the
associated 1D problems that generate them. (51) explicitly shows that $%
V^{(k)}_{0}$ and $V^{(n)}_{2}$ are generated from $V_{1}$ by the Darboux
type transformations. The functions that generate these transformations are
the eigenfunctions of the associated 1D problems. This constitutes an
extension of Darboux transformations for 2D problems. Another point worth
emphasizing is that any solution of these 1D problems can be used in
generating the potentials. But, as easily accessible results from the
literature, only normalizable solutions of these problems will be presented
below. From now on we take $\phi=0$ and in Secs. VIII and X we include $%
2m/\hbar^{2}$ into our notation.

\section{Bound states of the Associated Problems and $V_{1}$}

Provided that $c\geq -1/4$, the bound states of $H_{SO}$ belonging to the
Hilbert space $L^{2}(0,\infty )$ are given as follows \cite
{Fris,Perelomov,Lathouvers,Fuchs} 
\begin{eqnarray}
\Psi _{n}^{\varepsilon }(u) &=&N_{n}^{SO}u^{\frac{1}{2}+\varepsilon \nu
}e^{-\beta ^{2}u^{2}/2}L_{n}^{\varepsilon \nu }(\beta ^{2}u^{2}),  \nonumber
\\
E_{n}^{\varepsilon } &=&\frac{\hbar ^{2}}{2m}\lambda _{2,n}^{\varepsilon
}=\hbar \omega (2n+\varepsilon \nu +1),\quad n=0,1,2,..., \\
N_{n}^{SO} &=&[\frac{n!2\beta ^{2(1+\varepsilon \nu )}}{\Gamma
(n+\varepsilon \nu +1)}]^{1/2},\qquad \nu =\frac{1}{2}(1+4c)^{1/2}, 
\nonumber
\end{eqnarray}
where $N_{n}^{SO}$ is the normalization constant, $L_{n}^{\varepsilon
\nu}(z) $ are the generalized Laguerre polynomials, $\beta $ is defined by
Eq. (41), $\Gamma $ stands for the Gamma function and $\varepsilon =\pm $. $%
\Psi _{n}^{\varepsilon }(u)$'s satisfy the orthogonality relation \cite
{Magnus} 
\begin{equation}
\int_{0}^{\infty }\Psi _{n}^{\varepsilon }(u)\Psi _{n^{\prime
}}^{\varepsilon }(u)du=\delta _{nn^{\prime }},
\end{equation}
which is valid for $\varepsilon \nu >-1$. This implies that for $c\in
I=[-1/4,3/4)$ (that is for $-1/4\leq c<3/4$ ) both values of $%
\varepsilon=\pm $, and for $c\geq 3/4$ only $\varepsilon =+$ can be used for
each $n$. Although the generated potentials do not depend on the
normalization constants of the associated 1D problems we write them for
completeness.

From the most general point of view and in accordance with the fact that $%
H_{SO}$ is parity invariant, defined parity states of $H_{SO}$ belonging to
the Hilbert space $L^{2}(-\infty, \infty )$ can be given as follows \cite
{Lathouvers} 
\begin{eqnarray}
\Psi^{\varepsilon }_{n}(u) =\frac{1}{2^{1/2}}N^{SO}_{n} \left\{ 
\begin{array}{cc}
|u|^{\frac{1}{2}+\varepsilon \nu } e^{-\beta ^{2}u^{2}/2}L_{n}^{\varepsilon
\nu}(\beta ^{2}u^{2}); & \quad for \quad u\geq 0, \\ 
&  \\ 
-\varepsilon |u|^{\frac{1}{2}+\varepsilon \nu } e^{-\beta
^{2}u^{2}/2}L_{n}^{\varepsilon \nu}(\beta ^{2}u^{2}); & \quad for \quad u<0.
\end{array}
\right.
\end{eqnarray}
These obey the following orthogonality relation 
\begin{eqnarray}
\int_{-\infty}^{\infty}\Psi^{\varepsilon } _{n}(u) \Psi^{\bar{\varepsilon}}
_{n^{\prime}}(u) du= \delta_{n n^{\prime}} \delta_{\varepsilon \bar{%
\varepsilon}},
\end{eqnarray}
where $\varepsilon, \bar{\varepsilon}$ may equal $\pm$. For $\varepsilon= 
\bar{\varepsilon}$ (55) follows from the orthogonality of the generalized
Laguerre polynomials \cite{Magnus} and for $\varepsilon\neq \bar{\varepsilon}
$ from the parity reasons as can be verified directly from (54). The
corresponding energy eigenvalues are given by (52). For $c<-1/4$ the energy
spectrum is not bounded from below which implies ``falling of the particle
to the center" and physical interpretation is lost \cite{Lathouvers,Fuchs}.
As $c\rightarrow 0,\nu \rightarrow 1/2$ and $\Psi^{\varepsilon}_{n}(u)$'s go
over, for $\varepsilon=+ $ to odd parity and for $\varepsilon=-$ to even
parity harmonic oscillator wave functions. This follows from the relations
between the Hermite and Laguerre polynomials \cite{Lathouvers,Magnus}. The
corresponding limits of the energy eigenvalues are obvious from (52).

The normalized eigenfunctions and corresponding eigenvalues of $V_{1}$ can
now be written as 
\begin{eqnarray}
\Psi _{\ell }^{(1)\bar{\varepsilon} \varepsilon}(x,y) &=&\Psi _{n_{1}}^{\bar{%
\varepsilon}}(x)\Psi _{n_{2}}^{\varepsilon }(y),  \nonumber \\
E_{\ell }^{\bar{\varepsilon}\varepsilon } &=&E_{n_{1}}^{\bar{\varepsilon}%
}+E_{n_{2}}^{\varepsilon }=\hbar \omega (2\ell +\bar{\varepsilon}\bar{\nu}%
+\varepsilon \nu +1), \\
\ell &=&n_{1}+n_{2},\qquad \ell ,n_{1},n_{2}=0, 1, 2,...  \nonumber
\end{eqnarray}
where $\bar{\nu}=(1+4b)^{1/2}/2$. $\Psi _{n_{1}}^{\bar{\varepsilon}}(x),
\Psi _{n_{2}}^{\varepsilon }(y)$ and $E_{n_{1}}^{\bar{\varepsilon}%
},E_{n_{2}}^{\varepsilon }$ are given by (52) (or, (54)) with suitable
replacements of the parameters and variables. It follows that bound states
of $V_{1}$ exist for $b,c\geq -1/4$. For $b,c\in I$ there are four different
states for each value of $\ell $. In the case of $b\in I,c\geq 3/4$, or $%
c\in I,b\geq 3/4$ there are two different states for each value of $\ell $,
and one state in the case of $b,c\geq 3/4$. In each case, for a given value
of $\ell $ the state with energy $E_{\ell }^{\bar{\varepsilon}\varepsilon }$
is $(\ell +1)$-fold degenerate. We should also note that if we require the
wavefunctions to be separately continuous at the origin the interval $%
I=[-1/4,3/4)$ and the conditions $b,c\geq 3/4$ must be replaced as $%
I=[-1/4,0]$ and as $b,c\geq 0$.

The singular oscillator problem has the spectrum generating algebra $%
su(1,1)=\{J_{0},J_{\pm }:[J_{0},J_{\pm }]=\pm J_{\pm
},[J_{+},J_{-}]=-2J_{0}\}$ realized as \cite{Perelomov,Letourneau} 
\begin{eqnarray}
J_{0}=\frac{H_{SO}}{4\beta ^{2}},\quad J_{\pm }=-\frac{1}{4}[\beta ^{2}(u\mp
\beta ^{-2}\partial _{u})^{2}-\frac{c}{\beta ^{2}u^{2}}],
\end{eqnarray}
with the Casimir invariant $C^{2}=-J_{+}J_{-}+J_{0}^{2}-J_{0}=(4c-3)/16$.
Therefore, as will be shown in the next section, the symmetry algebra of the 
$H_{1}$-problem is closely connected with this kind two commuting copies of $%
su(1,1)$ algebra.

For later use it will be convenient to consider $H_{PT}$-problem in relation
with the solution of $H_{1}$-problem in the polar coordinates. In this case
the eigenvalue equation of $H_{1}$ separates, by taking $\Psi
^{(1)}(r,\theta )=R_{k_{1}}(r)\psi_{k}(\theta )$, into the P\"{o}schl-Teller
problem given by (48) and into the radial equation 
\begin{eqnarray}
\lbrack -(\frac{d^{2}}{d\rho ^{2}}+\frac{1}{\rho }\frac{d}{d\rho })+\rho
^{2}+\frac{\lambda _{1,k}}{\rho ^{2}}]R_{k_{1}}(\rho )=\lambda
R_{k_{1}}(\rho ),
\end{eqnarray}
where $\rho =\beta r$ and $\lambda =E/\beta ^{2}$. In terms of $v=\sin
^{2}\theta$, and $\psi _{k}(\theta )= v^{\frac{1}{2}(\frac{1}{2}+\varepsilon
\nu )}(1-v)^{\frac{1}{2} (\frac{1}{2}+\bar{\varepsilon}\bar{\nu})}F(v)$, the
eigenvalue equation of $H_{PT}$ leads us to the hypergeometric equation for $%
F(v)$: 
\begin{eqnarray}
v(1-v)\frac{d^{2}F}{dv^{2}}+[\zeta -v(\gamma +\eta +1)]\frac{dF}{dv}-\gamma
\eta F=0.  \nonumber
\end{eqnarray}
The general solution of this equation is 
\begin{eqnarray}
F(v)=A_{2}F_{1}(\gamma ,\eta ;\zeta ;v)+Bv^{1-\zeta }{}_{2}F_{1}(\gamma
-\zeta +1,\eta -\zeta +1;2-\zeta ;v),  \nonumber
\end{eqnarray}
where $A$ and $B$ are arbitrary constants, $_{2}F_{1}$ denotes the
hypergeometric function and 
\begin{eqnarray}
\gamma =\frac{1}{2}(1+\varepsilon \nu +\bar{\varepsilon}\bar{\nu}+\surd
\lambda _{1,k}),\quad \eta =\frac{1}{2}(1+\varepsilon \nu +\bar{\varepsilon}%
\bar{\nu}-\surd \lambda _{1,k}),\quad \zeta =1+\varepsilon \nu .  \nonumber
\end{eqnarray}
For normalizable solutions $B$ must be zero and $\gamma $ (or $\eta $) must
be a negative integer, say, $-k$. In that case the hypergeometric function
goes over to Jacobi polynomials $P_{k}^{(\varepsilon \nu ,\bar {\varepsilon}%
\bar{\nu})}(1-2v)$ and the resulting eigenfunctions and eigenvalues can be
written as follows \cite{Evans2,Flugge} 
\begin{eqnarray}
\psi _{k}(\theta ) &=&N_{k}^{PT}\sin ^{\frac{1}{2}+\varepsilon \nu }\theta
\cos ^{\frac{1}{2}+\bar{\varepsilon }\bar{\nu}}\theta P_{k}^{(\varepsilon
\nu, \bar{\varepsilon}\bar{\nu})}(\cos 2\theta ),  \nonumber \\
E_{k} &=&\frac{\hbar ^{2}}{2m}\lambda _{1,k}=\frac{\hbar ^{2}}{2m}%
(2k+\varepsilon \nu +\bar{\varepsilon }\bar{\nu}+1)^{2},\qquad k=0,1,2,...,
\\
N^{PT}_{k} &=&[\frac{2(2k+\varepsilon \nu +\bar{\varepsilon }\bar{\nu}%
+1)\Gamma (k+1)\Gamma (k+\varepsilon \nu +\bar{\varepsilon }\bar{\nu}+1)}{%
\Gamma (k+\varepsilon \nu +1)\Gamma (k+\bar{\varepsilon }\bar{\nu}+1)}%
]^{1/2}.  \nonumber
\end{eqnarray}

Substituting $\lambda _{1,k}=(2k+\varepsilon \nu +\bar{\varepsilon}\bar{\nu}%
+1)^{2}$ into Eq. (58) and trying the solution $R_{k_{1}}(\rho )=\rho ^{\mu
}e^{-\rho ^{2}/2}G_{k_{1}}(\rho )$ we end up, for $\mu =\surd \lambda
_{1,k}=(2k+\varepsilon \nu +\bar{\varepsilon}\bar{\nu}+1)$, with the
equation 
\begin{equation}
z\frac{d^{2}G_{k_{1}}}{dz^{2}}+(\mu +1-z)\frac{dG_{k_{1}}}{dz}-\frac{1}{4}%
[2(\mu +1)-\lambda ]G_{k_{1}}=0,
\end{equation}
where $z=\rho ^{2}$. Provided that $-[2(\mu +1)-\lambda ]/4$ is an integer,
say $k_{1}=0,1,2,...$, the solutions of (60) are the generalized Laguerre
polynomials. Hence, the radial solutions are 
\begin{eqnarray}
R_{k_{1}}(\rho )=N_{k_{1}}\rho ^{\mu }e^{-\rho ^{2}/2}L_{k_{1}}^{\mu }(\rho
^{2}),\quad N_{k_{1}}=[\frac{2\Gamma (k_{1}+1)}{\Gamma (k_{1}+\mu +1)}%
]^{1/2}.
\end{eqnarray}
One can easily verify that $\psi _{k}(\theta )$'s and $R_{k_{1}}(\rho )$'s
obey the following orthogonality relations 
\begin{eqnarray}
\int_{0}^{\infty }R_{k_{1}}(\rho )R_{k_{1}^{\prime }}(\rho )rdr=\delta
_{k_{1}k_{1}^{\prime }},\qquad \int_{0}^{\pi /2}\psi _{k}(\theta )\psi
_{k^{\prime }}(\theta )d\theta =\delta _{kk^{\prime }}.
\end{eqnarray}

As a result the eigenfunctions of $H_{1}$ can be written in polar
coordinates as follows 
\begin{equation}
\Psi _{\ell }^{(1)\bar{\varepsilon}\varepsilon }(r,\theta
)=N_{k_{1}}N_{k_{2}}^{PT}(\beta r)^{\mu }e^{-\beta ^{2}r^{2}/2}\sin ^{\frac{1%
}{2}+\varepsilon \nu }\theta \cos ^{\frac{1}{2}+\bar{\varepsilon}\bar{\nu}%
}\theta L_{k_{1}}^{\mu }(\beta ^{2}r^{2})P_{k_{2}}^{(\varepsilon \nu ,\bar{%
\varepsilon}\bar{\nu})}(\cos 2\theta ),
\end{equation}
with $\ell =k_{1}+k_{2};\ell ,k_{1},k_{2}=0,1,2,...$, and $\mu
=(2k_{2}+\varepsilon \nu +\bar{\varepsilon}\bar{\nu}+1)$. Since $\psi
_{k}(\theta )$ given by (59) will be used in generating $V_{0}^{(k)}$
potentials, in writing (63) we have changed the quantum number $k$ as $k_{2}$%
. Observe that a similar change ($n\rightarrow n_{2}$) has been made in
writing (56). Note also that the condition $-[2(\mu +1)-\lambda ]/4=k_{1}$
gives the eigenvalue (56) for the $V_{1}$-problem, with $\ell =k_{1}+k_{2}$.
We should also note that, as has been done in Eq. (54), the solutions (63)
may be extended to all $xy$-plane such that they have definite parity under
2D parity transformation: $(r,\theta )\rightarrow (r,\theta +\pi )$.

Inserting $\psi _{k}$ and $\Psi _{n}$ into (51) explicit expressions of the
potentials labelled by the quantum numbers $k$ and $n$ become available.
Besides that presented in the Sec. VI several more special subfamilies can
be identified. In doing that one should take care of the range of the
parameters and the domain of definition for potentials. The bound states of $%
V_{0}^{(k)}$ and $V_{2}^{(n)}$ will be taken up in Sec. X after considering
the symmetry generators in the next section.

\section{Symmetry Generators and Their Algebras}

As is apparent from previous two sections, the intertwining operators,
symmetry generators, and the Hamiltonians $H_{0},H_{2}$ must be labelled by
the quantum numbers ($k ,n$) of the associated potentials. In terms of $e(2)$
generators the labelled intertwining operators are 
\begin{eqnarray}
{\cal L}^{(k)}_{10}&=&f_{k}(\theta)+J,\qquad {\cal L}^{(k)%
\dagger}_{10}=f_{k} (\theta)-J,  \nonumber \\
{\cal L}^{(n)} _{21}&=&g_{n} (u)+ T_{2},\quad {\cal L}^{(n) \dagger}_{21}=
g_{n} (u)-T_{2}.
\end{eqnarray}
It is easy to verify that they obey the following commutators 
\begin{eqnarray}
\lbrack{\cal L}^{(k)} _{10}, {\cal L}_{10}^{(k) \dagger}]&=& 2f_{k}
^{\prime}(\theta),\qquad \quad \lbrack{\cal L}^{(n)} _{21}, {\cal L}%
_{21}^{(n)\dagger}]=2g_{n} ^{\prime}(u),  \nonumber \\
\lbrack{\cal L}^{(k)} _{10}, {\cal L}^{(n)} _{21}]&=&K^{(k,n)}
_{-}+T_{1},\quad \lbrack{\cal L}^{(k) \dagger}_{10}, {\cal L}%
_{21}^{(n)\dagger}]=-K^{(k,n)} _{-}+T_{1}, \\
\lbrack{\cal L}^{(k)}_{10}, {\cal L}^{(n) \dagger}_{21}]&=&K^{(k,n)}
_{+}-T_{1},\quad \lbrack{\cal L}^{(k)\dagger}_{10}, {\cal L}^{(n)}
_{21}]=-K^{(k,n)} _{+}-T_{1},  \nonumber
\end{eqnarray}
where 
\begin{eqnarray}
K^{(k,n)} _{\pm}=r\cos\theta[g_{n}^{\prime}(u)\pm \frac{1}{r^{2}}f_{k}
^{\prime}(\theta)].  \nonumber
\end{eqnarray}
By virtue of (28) we have 
\begin{eqnarray}
K^{(k,n)}_{+}=\frac{1}{2}r\cos\theta[H_{2}^{(k)}-H_{0}^{(n)}],\qquad
K^{(k,n)}_{-}=\frac{1}{2}r\cos\theta[H_{0}^{(k)}+H_{2}^{(n)}-2H_{1}].
\end{eqnarray}

It will be convenient to start with the symmetry generators of $H_{1}$ 
\begin{eqnarray}
X^{(k)} _{1} &=&{\cal L}^{(k)} _{10}{\cal L}_{10}^{(k)\dagger
}=H_{PT}-\lambda _{1,k}, \\
Y^{(n)} _{1} &=&{\cal L}_{21}^{(n)\dagger }{\cal L}^{(n)}
_{21}=H_{SO}-\lambda _{2,n},
\end{eqnarray}
where we have made use of $T_{2}g_{n} (u)=g_{n} ^{\prime }(u)$. $H_{PT}$ and 
$H_{SO}$ are defined by (49) and (50). Second order symmetry generators of $%
H^{(k)}_{0}$ and $H^{(n)}_{2}$ can also be written as follows 
\begin{eqnarray}
X^{(k)}_{0}&=&{\cal L}^{(k)\dagger}_{10}{\cal L}^{(k)}_{10}=
V^{(k)}_{-}-J^{2}=\bar{H}^{(k)}_{PT}-\lambda_{1,k}, \\
X^{(n)}_{2}&=&{\cal L}^{(n)}_{21}{\cal L}^{(n)\dagger} _{21}= g_{n}
^{2}+g_{n}^{\prime}-T^{2}_{2}= \bar{H}^{(n)} _{SO}-\lambda_{2,n},
\end{eqnarray}
where 
\begin{eqnarray}
\bar{H}^{(k)} _{PT}&=& -\frac{d^{2}}{d\theta^{2}}+V_{PT}-
2\partial^{2}_{\theta}\ln \psi_{k} (\theta), \\
\bar{H}^{(n)}_{SO}&=&-\frac{d^{2}}{du^{2}}+V_{SO}- 2\partial^{2}_{u}\ln
\Psi_{n} (u).
\end{eqnarray}
These are the so called super partners of $H_{PT}$ and $H_{SO}$. As a
result, the Hamiltonians of 1D auxiliary problems are, up to some constants,
the second order symmetry generators of $H_{1}$ and their super partners are
the second order symmetry generators of $H^{(k)}_{0}$ and $H^{(n)}_{2}$.

The simplest forms of remaining fourth order generators seem to be their
factorized forms given by (5). Making use of (65) and (67-70) these can be
expressed in a variety of ways, some of which are as follows; 
\begin{eqnarray}
Y_{0}^{(k,n)} &=&{\cal L}_{10}^{(k)\dagger }Y_{1}^{(n)}{\cal L}_{10}^{(k)}=%
{\cal L}_{10}^{(k)\dagger }H_{SO}{\cal L}_{10}^{(k)}-\lambda
_{2,n}X_{0}^{(k)},  \nonumber \\
&=&Y_{1}^{(n)}X_{0}^{(k)}-[(K_{-}^{(k,n)}-T_{1}){\cal L}_{21}^{(n)}+{\cal L}%
_{21}^{(n)\dagger }(K_{+}^{(k,n)}+T_{1})]{\cal L}_{10}^{(k)}, \\
&=&X_{0}^{(k)}Y_{1}^{(n)}-{\cal L}_{10}^{(k)\dagger }[(K_{+}^{(k,n)}-T_{1})%
{\cal L}_{21}^{(n)}+{\cal L}_{21}^{(n)\dagger }(K_{-}^{(k,n)}+T_{1})], 
\nonumber \\
Y_{2}^{(k,n)} &=&{\cal L}_{21}^{(n)}X_{1}^{(k)}{\cal L}_{21}^{(n)\dagger }=%
{\cal L}_{21}^{(n)}H_{PT}{\cal L}_{21}^{(n)\dagger }-\lambda
_{1,k}X_{2}^{(n)},  \nonumber \\
&=&X_{1}^{(k)}X_{2}^{(n)}-[(K_{-}^{(k,n)}+T_{1}){\cal L}_{10}^{(k)\dagger }-%
{\cal L}_{10}^{(k)}(K_{+}^{(k,n)}+T_{1})]{\cal L}_{21}^{(n)\dagger }, \\
&=&X_{2}^{(n)}X_{1}^{(k)}-{\cal L}_{21}^{(n)}[{\cal L}%
_{10}^{(k)}(K_{-}^{(k,n)}-T_{1})-(K_{+}^{(k,n)}-T_{1}){\cal L}%
_{10}^{(k)\dagger }].  \nonumber
\end{eqnarray}
At this point we have to emphasize the followings. The existence of $%
\lambda_{1,k}$ and $\lambda _{2,n}$ as additive constants in $X_{1}^{(k)},
Y_{1}^{(n)}$ and $X_{0}^{(k)}, X_{2}^{(n)}$ seems to be redundant in regard
of superintegrability of $H_{1}$. In particular, our labelling of the fourth
order generators with two indices may seem as if we have more symmetries
than is needed for superintegrability. But an inspection of the first lines
of Eqs. (73) and (74) immediately shows that, for a given, say, $k$ and all $%
n$ the set $\{Y_{0}^{(k,n)},X_{0}^{(k)}\}$ spans only a 2D vector space. As
fourth order symmetries labelled with one index one may take 
\begin{eqnarray}
\bar{Y}_{0}^{(k)} &\equiv &Y_{0}^{(k,n)}+\lambda _{2,n}X_{0}^{(k)}={\cal L}%
_{10}^{(k)\dagger }H_{SO}{\cal L}_{10}^{(k)},  \nonumber \\
\bar{Y}_{2}^{(n)} &\equiv &Y_{2}^{(k,n)}+\lambda _{1,k}X_{2}^{(n)}={\cal L}%
_{21}^{(n)}H_{PT}{\cal L}_{21}^{(n)\dagger }-\lambda _{1,k}X_{2}^{(n)}. 
\nonumber
\end{eqnarray}
However, for overall consistency of the hierarchy such as intertwining of $%
H_{1}$ with $H_{0}^{(k)}, H_{2}^{(n)}$ and, as we will show in the next
section, in determining the spectra of $H_{0}^{(k)}$ and $H_{2}^{(n)}$ these
seemingly redundant constants and labels play an essential role.

One of virtues of our approach is that the commutativity of the symmetry
generators with the corresponding Hamiltonian is guaranteed by construction
from the outset. For justification we first note that the relations 
\begin{eqnarray}
\lbrack H_{1},X_{1}^{(k)}]=0=[H_{1},Y_{1}^{(k)}],\qquad k=0,1,...,
\end{eqnarray}
immediately follow from the fact that $X_{1}^{(k)}$ and $Y_{1}^{(k)}$ emerge
from the separation of $H_{1}$ in different coordinate systems. Secondly at
a glimpse of Eqs. (67-72) we observe that 
\[
X_{0}^{(k)}=X_{1}^{(k)}-2\partial _{\theta }^{2}\ln \psi _{k}(\theta
),\qquad X_{2}^{(n)}=Y_{1}^{(n)}-2\partial _{u}^{2}\ln \Psi _{n}(u). 
\]
That is, the second order symmetry generators of $H_{0}^{(k)}$ and $%
H_{2}^{(n)}$ are Darboux transforms of symmetry generators of $H_{1}$ as
are, apart from the factor $r^{-2}$, $H_{0}^{(k)}$ and $H_{2}^{(n)}$ Darboux
transforms of $H_{1}$ along different legs of diagram (3). In view of this
fact the relations 
\begin{eqnarray}
\lbrack H_{0}^{(k)},X_{0}^{(k)}]=0=[H_{2}^{(n)},X_{2}^{(n)}],\qquad
n,k=0,1,...,
\end{eqnarray}
follow from, or, in a sense, are Darboux transforms of (75). Only the
explicit check of 
\begin{eqnarray}
\lbrack H_{0}^{(k)},Y_{0}^{(k,n)}]=0=[H_{2}^{(n)},Y_{2}^{(k,n)}],\qquad
n,k=0,1,...,
\end{eqnarray}
takes tediously a lot of time. This shows an advantage of our method
compared with the conventional approach in which much effort is devoted to
verify the commutativity for specified forms of generators. There it is
known that for symmetries higher than second order, equations resulting from
commutativity are almost intractable.

It is not so hard to check that $\lbrack X^{( )}_{j}, Y^{( )}_{j}]\neq 0,
j=0, 1, 2,$ since the highest order derivatives with constant coefficients
will appear at the right hand side. For example, 
\begin{eqnarray}
\lbrack X^{(k)} _{0}, Y^{(k,n)} _{0}]_{hot}&=&[J^{2},JT_{2}^{2}J]=
4T_{1}T_{2}J^{3}+2(2T^{2}_{1}-2T^{2}_{2}+T_{1}T_{2})J^{2}-8T_{1}T_{2}J , 
\nonumber \\
\lbrack X^{(n)} _{2}, Y^{(k,n)} _{2}]_{hot}&=& [T^{2}_{2},T_{2}J^{2}T_{2}]=
2T_{2}^{4}-6T^{2}_{1}T^{2}_{2}-4T_{1}T_{2}^{3}J,  \nonumber
\end{eqnarray}
where $[X^{(k,n)} _{0}, Y^{(k,n)} _{0}]_{hot}$ represents only the highest
order terms resulting from $[X^{(k)} _{0}, Y^{(k,n)} _{0}]$. Therefore, the
symmetry generators of each potential do not close in a finite dimensional
Lie algebra structure. Note that by the Jacobi identity $%
Z_{j}=[X_{j},Y_{j}],j=0,1,2,$ is also a symmetry generator, but it is
algebraically dependent to $X_{j}$ and $Y_{j}$.

There is an elegant way of expressing the symmetry algebra of $H_{1}$. For
this purpose we introduce the generators 
\begin{eqnarray}
X_{\pm } &=&\frac{1}{4\beta ^{2}}(-\partial _{x}^{2}\pm \frac{1}{2}\alpha
x^{2}+\frac{b}{x^{2}}),\quad D_{1}=\frac{1}{4}(1+ 2x\partial _{x}), \\
Y_{\pm } &=&\frac{1}{4\beta ^{2}}(-\partial _{y}^{2}\pm \frac{1}{2}\alpha
y^{2}+\frac{c}{y^{2}}),\quad D_{2}=\frac{1}{4}(1+2y\partial _{y}),
\end{eqnarray}
which obey the Lie algebras 
\begin{eqnarray}
\lbrack X_{\pm },D_{1}] &=&X_{\mp },\quad \lbrack X_{+},X_{-}]=D_{1}, \\
\lbrack Y_{\pm },D_{2}] &=&Y_{\mp },\quad \lbrack Y_{+},Y_{-}]=D_{2},
\end{eqnarray}
with the Casimir invariants 
\begin{eqnarray}
X_{+}^{2}-X_{-}^{2}+D_{1}^{2}=\frac{4b-3}{16},\quad
Y_{+}^{2}-Y_{-}^{2}+D_{2}^{2}=\frac{4c-3}{16}.
\end{eqnarray}
It is straightforward to show that in terms of (78) and (79) we have 
\begin{eqnarray}
H_{1} &=&4\beta ^{2}(X_{+}+Y_{+}),  \nonumber \\
X_{1}^{(k)} &=&8(X_{+}Y_{+}-X_{-}Y_{-}+D_{1}D_{2})+K, \\
Y_{1}^{(n)} &=&4\beta ^{2}Y_{+}-\lambda _{2,n},  \nonumber
\end{eqnarray}
where $K=b+c-\lambda _{1,k}-(1/2)$. Eqs. (80-82) are defining relations of
two commuting copies of a $su(1,1)$ algebra which can be written as a direct
sum $su(1,1)\oplus su(1,1)$. The basis given by (78-79) is connected with
that mentioned in Sec. VIII by linear transformations, for instance, by
comparing (57) and (79) we have $Y_{+}=J_{0},Y_{-}\pm D_{2}=J_{\pm }$. Eqs.
(83) show that the symmetries of $H_{1}$ are quadratic in the generators of
centrally extended (because of the constant $K$) $su(1,1)\oplus su(1,1)$
algebra.

By defining 
\begin{eqnarray}
W^{(k,n)}\equiv \frac{1}{8}[X_{1}^{(k)},Y_{1}^{(n)}]=4\beta
^{2}(X_{-}D_{2}-Y_{-}D_{1}),
\end{eqnarray}
one can easily show that 
\begin{eqnarray}
\lbrack X_{1}^{(k)},W^{(k,n)}]
&=&\{X_{1}^{(k)},Y_{1}^{(n)}\}+X_{1}^{(k)}(2\lambda _{2,n}-H_{1})+  \nonumber
\\
& &(2Y_{1}^{(n)}+2\lambda _{2,n}-H_{1})(\lambda _{1,k}-1)+H_{1}(b-c), \\
\lbrack Y_{1}^{(n)},W^{(k,n)}] &=&(Y_{1}^{(n)}+\lambda
_{2,n})(H_{1}-Y_{1}^{(n)}-\lambda _{2,n})-2\beta^{4}(X_{1}^{(k)}-K),
\end{eqnarray}
where $\{,\}$ represents the anti commutator. These explicitly show that the
extended symmetry algebra of $H_{1}$ spanned by $%
\{H_{1},X_{1}^{(k)},Y_{1}^{(n)},W^{(k,n)}\}$, with the inclusion of $%
W^{(k,n)}$, closes in a quadratic associative algebra for all values of $k,n$%
. We also observe that this algebra is a cubic associative algebra in the
enveloping algebra of the centrally extended $su(1,1)\oplus su(1,1)$.
Recently such finitely generated associative algebras have attracted a great
deal of interest. The structure we have obtained coincides, up to some
additive constants, with that presented in Ref.\cite{Daskal} for the
Winternitz potential $V_{1}$. In Ref.\cite{Letourneau} this structure is
constructed as a cubic associative algebra in which counterparts of $%
X_{1},Y_{1}$ are taken to be purely quadratic in the generators of $%
su(1,1)\oplus su(1,1)$. We end this section by emphasizing that exploring
similar algebraic structures for $H_{0}^{(k)}$ and $H_{2}^{(n)}$ and
connection between them seems to be an important problem which deserves to
be taken up in another study.

\section{Bound States of $H_{0}^{(k)},H_{2}^{(n)}$ and Their Degeneracies}

Representing $\Psi _{n_{1}}^{\bar{\varepsilon}},\Psi _{n_{2}}^{\varepsilon }$%
, and $\Psi _{\ell }^{(1)\bar{\varepsilon}\varepsilon }$ given by (52) and
(56), in the Dirac notation, respectively by the kets $|n_{1}\bar{\varepsilon%
}>,|n_{2}\varepsilon >$ and $|1;\ell \bar{\varepsilon}\varepsilon >$, we
write (56) as follows 
\begin{eqnarray}
|1;\ell \bar{\varepsilon}\varepsilon >=|n_{1}\bar{\varepsilon}%
>|n_{2}\varepsilon >,\qquad \ell =n_{1}+n_{2}.
\end{eqnarray}
In this notation, the corresponding isospectral states of $H_{2}^{(n)}$ are 
\begin{eqnarray}
|2n;\ell \bar{\varepsilon}\varepsilon >={\cal L}_{21}^{(n)}|1;\ell \bar{%
\varepsilon}\varepsilon >.
\end{eqnarray}
From (53) (or, 55), (56), (68) and (87) one can easily show that 
\begin{eqnarray}
<2n;\ell \bar{\varepsilon}\varepsilon |2n;\ell \bar{\varepsilon}\varepsilon
> &=&<1;\ell \bar{\varepsilon}\varepsilon |Y_{1}^{(n)}|1;\ell \bar{%
\varepsilon}\varepsilon >,  \nonumber \\
&=&<n_{2}\varepsilon |H_{SO}|n_{2}\varepsilon >-\lambda _{2,n}, \\
&=&2\hbar \omega (n_{2}-n),  \nonumber
\end{eqnarray}
where $<\cdot |\cdot >$ represents the usual inner product of ${\cal H}%
=L^{2}(R^{2})$ and in the third line we have included $2m/\hbar ^{2}$ into
the notation. Since $\ell =n_{1}+n_{2}$, this implies that as physically
acceptable states only those with $\ell >n$ will survive in the spectrum of $%
H_{2}^{(n)}$. Moreover, the degeneracies of the survived states will be
shifted to $\ell -n$ since the states corresponding to $n_{2}\leq n$ can not
be normalized. As a result, the normalized states of $H_{2}^{(n)}$ are as
follows 
\begin{eqnarray}
|2n;\ell \bar{\varepsilon}\varepsilon >=[2\hbar \omega (n_{2}-n)]^{-1/2}{\cal %
L}_{21}^{(n)}|1;\ell \bar{\varepsilon}\varepsilon >,
\end{eqnarray}
provided that $\ell =n_{1}+n_{2}$ and $n_{2}>n$.

In a similar way, if we represent $\psi _{k_{2}}$ and $R_{k_{1}}$'s given by
(59) and (61), respectively by the kets $|k_{2}\bar{\varepsilon }\varepsilon
>$ and $|k_{1}\bar{\varepsilon}\varepsilon > $, the states given by (63) can
be expressed as 
\begin{eqnarray}
|1;\ell \bar{\varepsilon } \varepsilon >= |k _{1} \bar{\varepsilon }%
\varepsilon > |k_{2}\bar{\varepsilon } \varepsilon >,\qquad \ell =k
_{1}+k_{2}.
\end{eqnarray}
In that case the corresponding isospectral states of $H_{0}^{(k)}$ are $%
|0k;\ell \bar{\varepsilon } \varepsilon >= {\cal L}_{10}^{(k)\dagger}
|1;\ell \bar{\varepsilon } \varepsilon >$ and by virtue of (62), (67) and
(91) we have 
\begin{eqnarray}
<0k;\ell \bar{\varepsilon } \varepsilon |0k;\ell \bar{\varepsilon }
\varepsilon >&=& <1;\ell \bar{\varepsilon } \varepsilon |X^{(k)}_{1}|1;\ell 
\bar{\varepsilon } \varepsilon >,  \nonumber \\
&=&<k _{2}\bar{\varepsilon }\varepsilon |H_{PT}|k _{2} \bar{\varepsilon }%
\varepsilon >-\lambda _{1,k}, \\
&=&\frac{2\hbar^{2}}{m} (k_{2}-k)(k+k_{2}+ \bar{\varepsilon }\bar{\nu}
+\varepsilon \nu+1).  \nonumber
\end{eqnarray}
Hence, the normalized states of $H^{(k)}_{0}$ are 
\begin{eqnarray}
|0k;\ell \bar{\varepsilon } \varepsilon >= [\frac{2\hbar^{2}}{m}
(k_{2}-k)(k+k_{2}+ \bar{\varepsilon }\bar{\nu} +\varepsilon \nu+1)]^{-1/2} 
{\cal L}_{10}^{(k)\dagger}|1;\ell \bar{\varepsilon } \varepsilon >,
\end{eqnarray}
provided that $\ell =k_{1}+k_{2}$ and $k _{2}>k$. In this case the
degeneracy of the state $|0k;\ell \bar{\varepsilon } \varepsilon >$ is $%
\ell-k$. Explicit functional realizations of the states (91) and (93) can
easily be obtained by applying ${\cal L}_{21}^{(n)}, {\cal L}%
_{10}^{(k)\dagger}$ to the wave functions given by (56) and (63).

\section{CONCLUDING REMARKS}

The method of intertwining is a unified approach widely used in various
fields of physics and mathematics such as in investigating particle
propagation on a curved space \cite{Anderson2,Miller,Veselov}, in
constructing matrix-Hamiltonian to realize higher dimensional superalgebras 
\cite{Cannata,Andrianov}, in solving both ordinary and partial differential
equations \cite{Anderson2}, in generating exact solutions of non-stationary
Schr\"{o}dinger equation \cite{Cannata,Samsonov}, and in constructing
isospectral potentials in an arbitrary space dimension \cite{Kuru}. The
method we have introduced increases the power and enlarges the range of
applicability of the intertwining operator idea. It allows us to perform
Darboux transformations in higher dimensions in such a manner that, in
addition to their isospectral deformation property they acquire
integrability and superintegrability preserving property. In particular, as
we have shown the realization of this method for 2D systems generates two
infinite families of isospectral and superintegrable quantum systems
intertwined to a 2D Winternitz system. Work on 3D realization of the method
is in progress.

The space of purely second order operators quadratic in the generators of $%
e(2)$ has, under the adjoint action of $E(2)$, only four orbits whose
representatives can be taken to be; $T_{1}^{2},J^{2},J^{2}+a_{0}T_{1}^{2}$
and $T_{1}J+JT_{1}$, where $a_{0}$ is a constant. Existence of only four
types Winternitz potentials is closely connected with this orbit structure
since each corresponds to a different 2D orthogonal coordinate system \cite
{Fris,Miller}. $T_{1}^{2},J^{2}$ constitute the differential parts of the
symmetry generators of $V_{1}$ and account for its separation in the
Cartesian and polar (hence in elliptic) coordinates. Therefore, the
appearance of the Winternitz potential $V_{1}$ as the common member of two
families is of no surprise; it is a direct result of our orbit prescription
in constructing the intertwinings in Sec. V. We also observe that since only 
$T_{1}^{2}$ and $J^{2}$ can be factorized as ${\cal L}{\cal L}^{\dagger }$
(or, as ${\cal L}^{\dagger }{\cal L}$) the other three Winternitz potentials
can not be utilized as $V_{1}$ in the context of this paper. In this regard,
a combination of our method and the conventional approach may be used for
similar purposes. Finally we point out that what made it possible to
implement Darboux transformations in our approach is that when the
eigenvalue equation of $V_{1}$ is separated in the Cartesian and polar
coordinates, at least one of the separated equation is of the
Schr\"{o}dinger type.

\acknowledgements

We thank A. U. Y\i lmazer for useful conversations. This work was supported
in part by the Scientific and Technical Research Council of Turkey
(T\"{U}B\.{I}TAK).

\end{document}